\documentclass[a4paper,
               ]{jacow}
\usepackage{pdfpages,multirow,ragged2e} %
\makeatletter%
	\ifboolexpr{bool{xetex}}
	 {\renewcommand{\Gin@extensions}{.pdf,%
	                    .png,.jpg,.bmp,.pict,.tif,.psd,.mac,.sga,.tga,.gif,%
	                    .eps,.ps,%
	                    }}{}
\makeatother

\ifboolexpr{bool{xetex} or bool{luatex}} 
 {}                                      
 {\usepackage[utf8]{inputenc}}           

\usepackage[USenglish]{babel}
\usepackage{graphicx}
\usepackage{subcaption}
\ifboolexpr{bool{jacowbiblatex}}%
 {%
  \addbibresource{jacow-test.bib}
  \addbibresource{biblatex-examples.bib}
 }{}
\begin{document}

\title{Generation of high-power attosecond x-ray FEL pulses carrying orbital angular momentum}

\author{Chenzhi Xu\textsuperscript{1}, Shanghai Institute of Applied Physics, Chinese Academy of Sciences, Shanghai, China\\
        \textsuperscript{1}also at University of Chinese Academy of Sciences, Beijing, China\\
		Jiawei Yan\thanks{jiawei.yan@xfel.eu}, Gianluca Geloni, Christoph Lechner, European XFEL, Schenefeld, Germany\\
		Haixiao Deng\thanks{denghx@sari.ac.cn}, Shanghai Advanced Research Institute, Chinese Academy of Sciences, Shanghai, China}
	
\maketitle

\begin{abstract}
   X-ray beams carrying orbital angular momentum (OAM), are emerging as a powerful tool to probe matter. Recently, a method called self-seeded FEL with OAM (SSOAM)  has been proposed to generate high-power x-ray OAM pulses, which places the traditional optical elements in the linear regime of the FEL amplification process before saturation to reduce the thermal load of the optical element. In this work, we propose to utilize the SSOAM scheme to produce attosecond x-ray vortices with high intensity. Numerical simulations demonstrate the x-ray OAM pulses with peak powers of more than one hundred gigawatts and a pulse duration of the order of hundred attoseconds can be achieved using the proposed method.
\end{abstract}

\section{INTRODUCTION}

Optical vortices carrying orbital angular momentum (OAM), characterized by a helical phase-front expressed as exp(\textit{il$\phi$}) where $\phi$ denotes the azimuthal coordinate and \textit{l} signifies the topological charge, have been the subject of intense interest. OAM light can be used in optical tweezers and optical spanner to trap and control particles \cite{ashkin1970acceleration,he1995direct}. Besides, OAM light has proven successful in achieving quantum storage across diverse systems \cite{ding2015quantum,zhou2015quantum} and optical communication \cite{barreiro2008beating}. 

As the application of OAM light at optical wavelengths matures, people begin to seek shorter wavelength OAM light to reveal new physical phenomena \cite{shen2019optical}.For example,  hard x-ray helical dichroism has been demonstrated on disordered samples \cite{rouxel2022hard}. X-ray OAM light is mainly provided by synchrotron radiation facilities. Meanwhile, with the development of ultrafast science \cite{orfanos2019attosecond} and the appearance of attosecond OAM light \cite{wang2019intense,xu2021isolated,dorney2019controlling}, attosecond pulses carrying OAM, mainly produced by high-order harmonic generation (HHG) sources \cite{wang2019intense,xu2021isolated,dorney2019controlling}, are used to study photoemission dynamics \cite{hernandez2015quantum,hedse2018applications} and generate light springs \cite{pariente2015spatio}. 

However, there are challenges to produce attosecond OAM light in the x-ray range. It is difficult for synchrotron radiation devices to produce light at the attosecond level, while it is challenging to achieve hard X-ray level wavelengths with HHG-produced light. Compared with HHG and synchrotron radiation sources, modern x-ray free-electron lasers (XFELs) stand out as a potent tool for generating x-ray pulses characterized by high brightness and pulse durations spanning from tens of femtoseconds to few hundred attoseconds. Harmonic radiation generated in a helical undulator carries orbital angular momentum, as confirmed both theoretically and experimentally \cite{hemsing2014first,bahrdt2013first}. Meanwhile, SSOAM, which involves using an optical element to introduce a helical phase to the radiation pulse in the linear regime, has demonstrated the ability to produce femtosecond-scale x-ray OAM light \cite{yan2023self}.

In this study, we focus on the generation of attosecond OAM light. We present a novel method for generating x-ray vortices characterized by high OAM purity, peak power, and short pulse duration in the as range. The innovative scheme is built upon SSOAM and achieves high power by combining  enhanced self-amplified spontaneous emission (ESASE)\cite{zholents2005method} with SSOAM. 

\section{ATTOSECOND OAM PULSE}

\begin{figure}[h]
        \setlength{\abovecaptionskip}{-0.4cm}
        \captionsetup[subfloat]{labelsep=none,format=plain,labelformat=empty}
	\begin{center}
		\subfloat[\label{fig:1a}]{			
			\includegraphics[width=8.5cm]{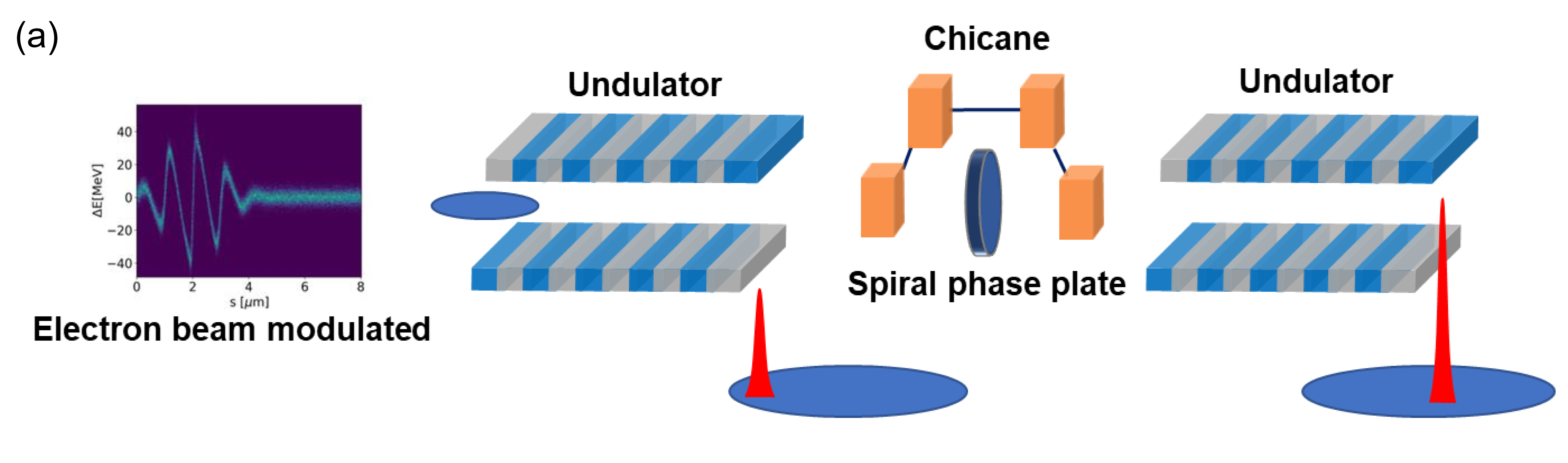}}
        \\
		\subfloat[\label{fig:1b}]{			
			\includegraphics[width=4cm]{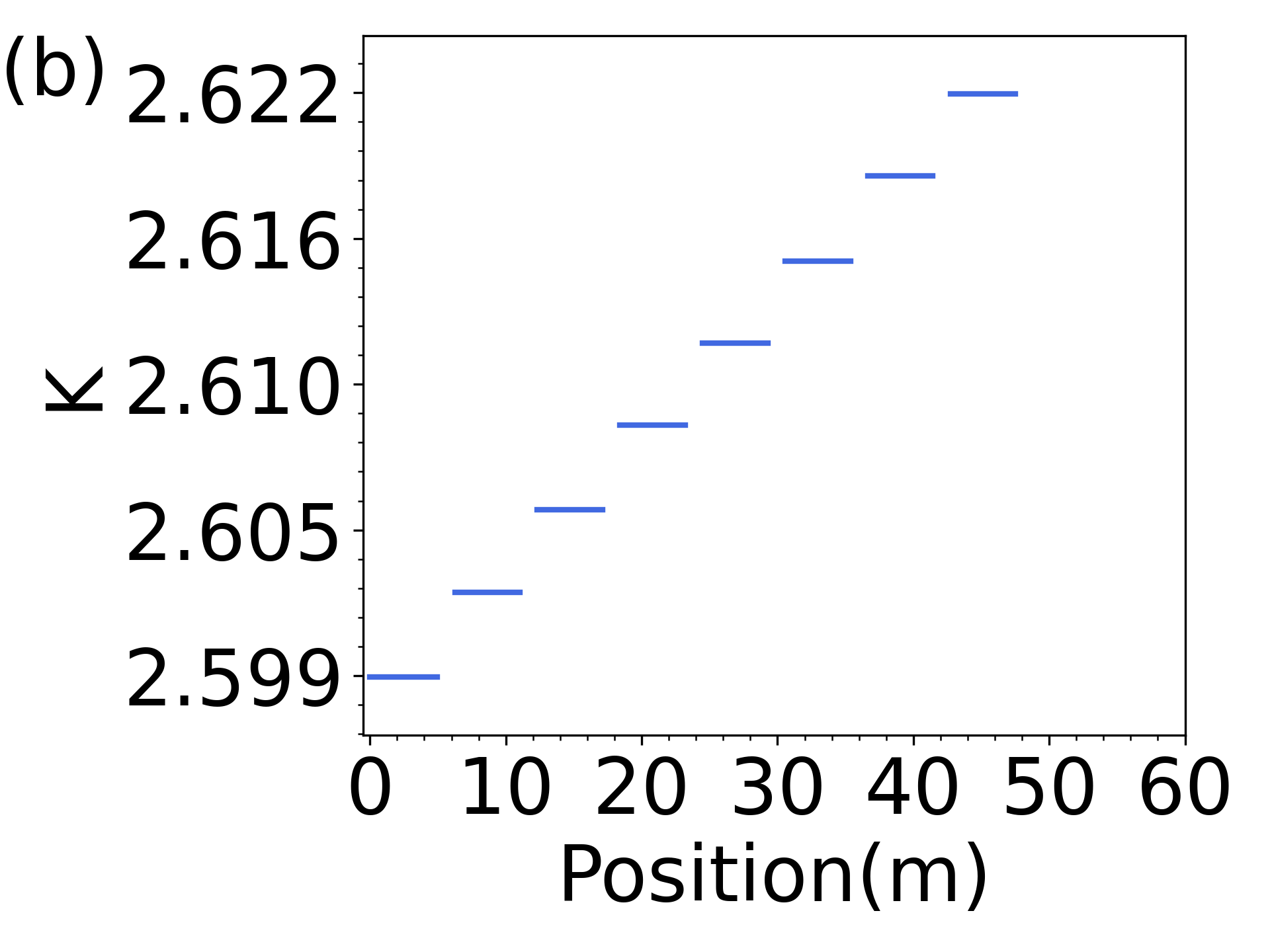}}
        \subfloat[\label{fig:1c}]{			
			\includegraphics[width=4cm]{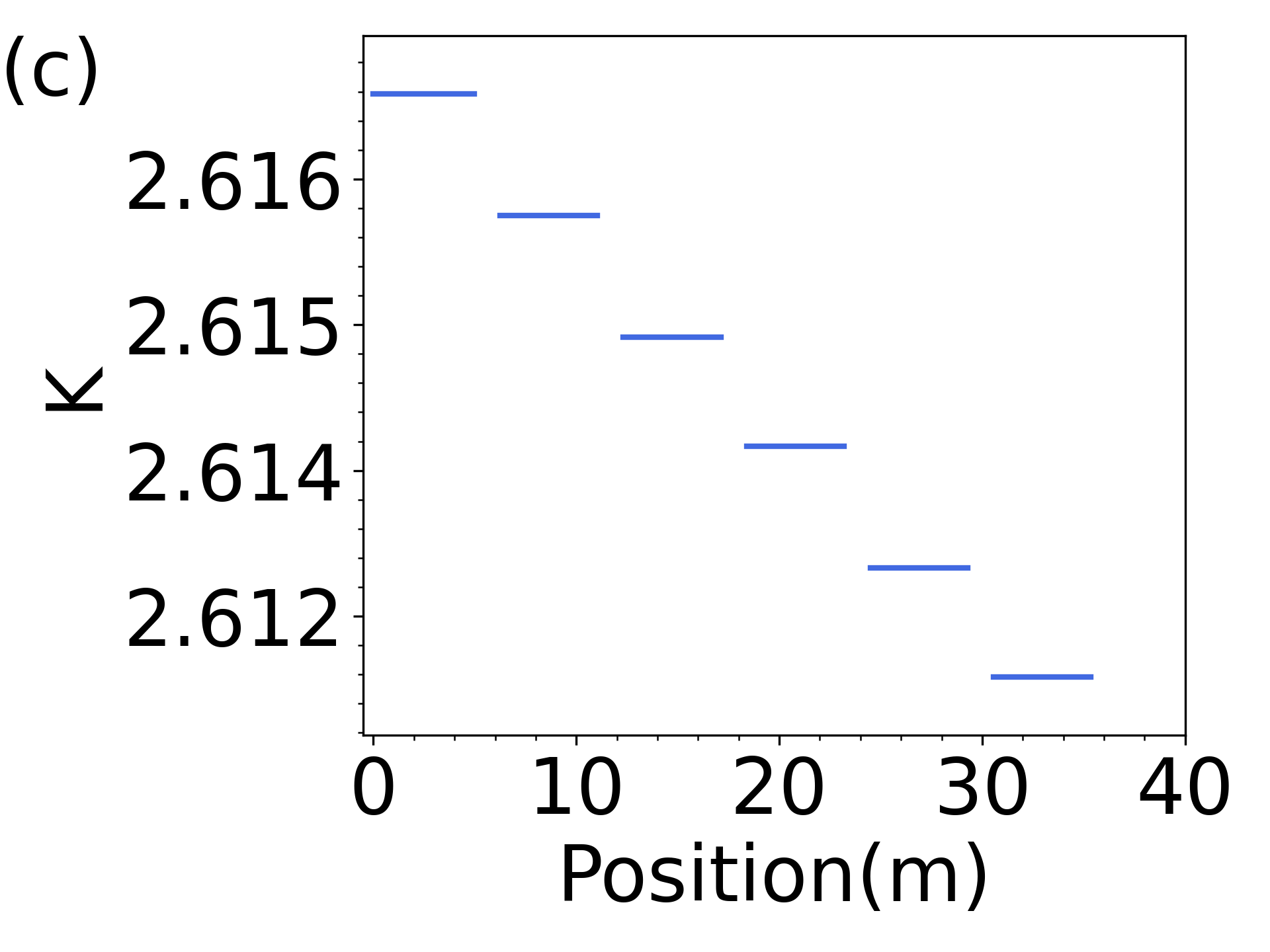}}
	\end{center}
	\caption{Schematic layout of the self-seeded FEL with attosecond OAM (a). Undulator parameter setting in the first stage (b) and the second stage (c).}
\end{figure}

The configuration of our proposed scheme is illustrated in Figure~\ref{fig:1a} assuming the presence of an ESASE setup preceding SSOAM. In other words, the electron beam initially passes through the linac and enters a wiggler magnet. At the same time a short fs laser enters the wiggler to modulate the electron beam in energy. Subsequently, the electron beam passes through a dispersive magnetic chicane to generate a periodic enhancement of the electron peak current. 

After modulation, the electron beam traverses a linearly tapered undulator, producing subfemtosecond x-ray pulses. The chirp-taper matching condition is expressed as \cite{duris2020superradiant, saldin2006self}:
\begin{equation}
	\begin{aligned}
		\frac{d K}{d z}=\frac{\lambda_r^2}{\lambda_u^2 K} \frac{d \gamma^2}{d s},
	\end{aligned}
        \label{eq2}
\end{equation}
where $\lambda_r$ represents the emission wavelength, $\lambda_u$ is the undulator period, K is the rms undulator parameter, and $\gamma$ is the Lorentz factor of the beam, z denotes the position along the undulator, and s represents the position along the electron bunch. Due to the nature of the chirp, lasing takes place solely in regions of the electron bunch where the chirp-taper matching condition is satisfied. Consequently, the modulation gives rise to a radiation pulse significantly shorter than the modulation period. As lasing is confined to regions meeting the chirp-taper matching condition in the electron bunch, a laser pulse with a duration in the order of a hundred attoseconds is generated in the tail of the electron beam.

To imprint the helical phase of a low-order OAM mode onto the FEL beam, we employ an optical element, such as a spiral phase plate (SPP) \cite{oemrawsingh2004production} or spiral Fresnel zone plate (SZP) \cite{vila2014characterization}. This process creates an OAM seed pulse with high purity. In order to host the optical element, the inclusion of a magnetic chicane is essential. This also destroys any residual microbunching from the first undulator. The primary objective is to precisely coincide the transformed XFEL beam with the unmodulated segment of the electron beam during the second stage of undulation, facilitating amplification. Therefore, it is imperative to employ a relatively long electron beam. 

The OAM seed pulse experiences substantial amplification within the second-stage undulator. Earlier investigations \cite{saldin2006self} delved into chirped-tapered FELs. After reaching the saturation point, extending the interaction involves tapering the undulator to sustain resonance as the radiation slips past the initial lasing region, facilitating interaction with fresh electrons. This process results in amplification of the initial pulse. 

The theoretical part of this scheme is same as the SSOAM \cite{yan2023self}. After the second stage, the light field can be represented as:
\begin{equation}
	\begin{aligned}
            \widetilde{E}(\hat{z}, \hat{r}, \phi)=&\sum_n A^{(n)} \sum_k \alpha_k \Phi_{n-l, k}^{(I I)}(\hat{r}) \\&\exp \left(\lambda_k^{(I I, n-l)} \hat{z}\right) \exp [-i(n-l) \phi]
        \end{aligned}
        \label{eq3}
\end{equation}
where $A^{(n)}$ is complex coupling constants, $\alpha_k$ are complex coefficients, $\Phi_{n-l, k}^{(I I)}$ are the FEL eigenfunctions of the second (II) stage, $\lambda_k^{(I I, n-l)}$ the corresponding eigenvalues, $\hat{r}=r / r_0$ and $\hat{z}=\Gamma z$ normalized versions of the physical longitudinal and radial coordinates $z$ and $r$ according to the definitions in \cite{saldin2000diffraction}, $n$ is azimuthal harmonic. In the second stage, amplification will be best for $m = n - l = 0$ and $k = 1$ that corresponds to the main FEL eigenmode of the second (II) stage (and zero topological charge). However, initially, the larger mode will be for $n = 0$ to the dominant contribution in the first (I) stage transformed to a topological charge of $h = l$ by the optical element and the following diffraction. In the initial part of the second stage, this topological charge is dominant (and will be amplified as well), but along the undulator it will lose the competition to the fundamental FEL mode of the second stage, which has zero topological charge and is amplified at a faster rate. Amplification must be stopped before that point.

\begin{figure}[h]
        \setlength{\abovecaptionskip}{-0.4cm}
        \captionsetup[subfloat]{labelsep=none,format=plain,labelformat=empty}
	\begin{center}
		\subfloat[\label{fig:2a}]{			
			\includegraphics[width=4cm]{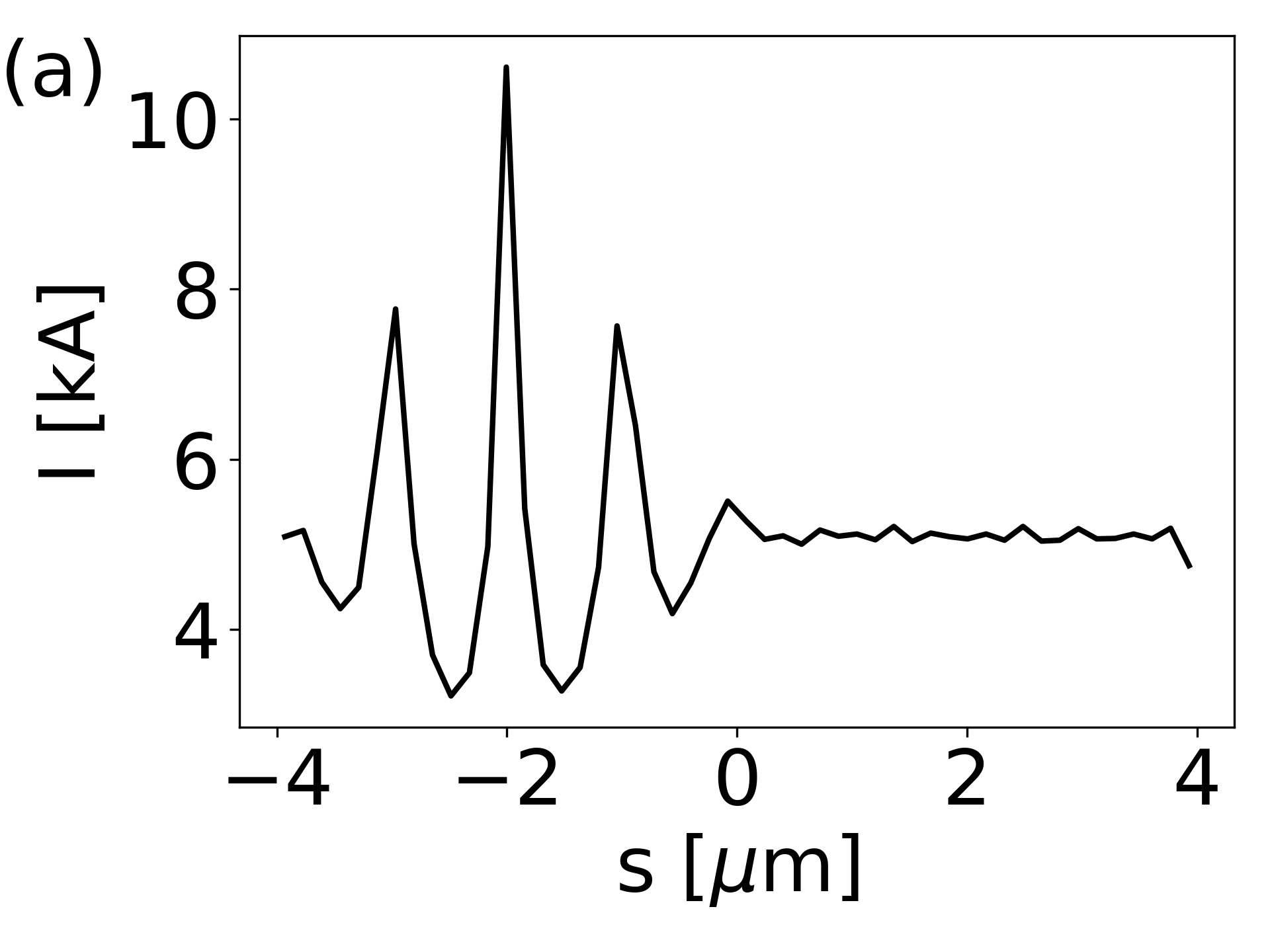}}
		\subfloat[\label{fig:2b}]{			
			\includegraphics[width=4cm]{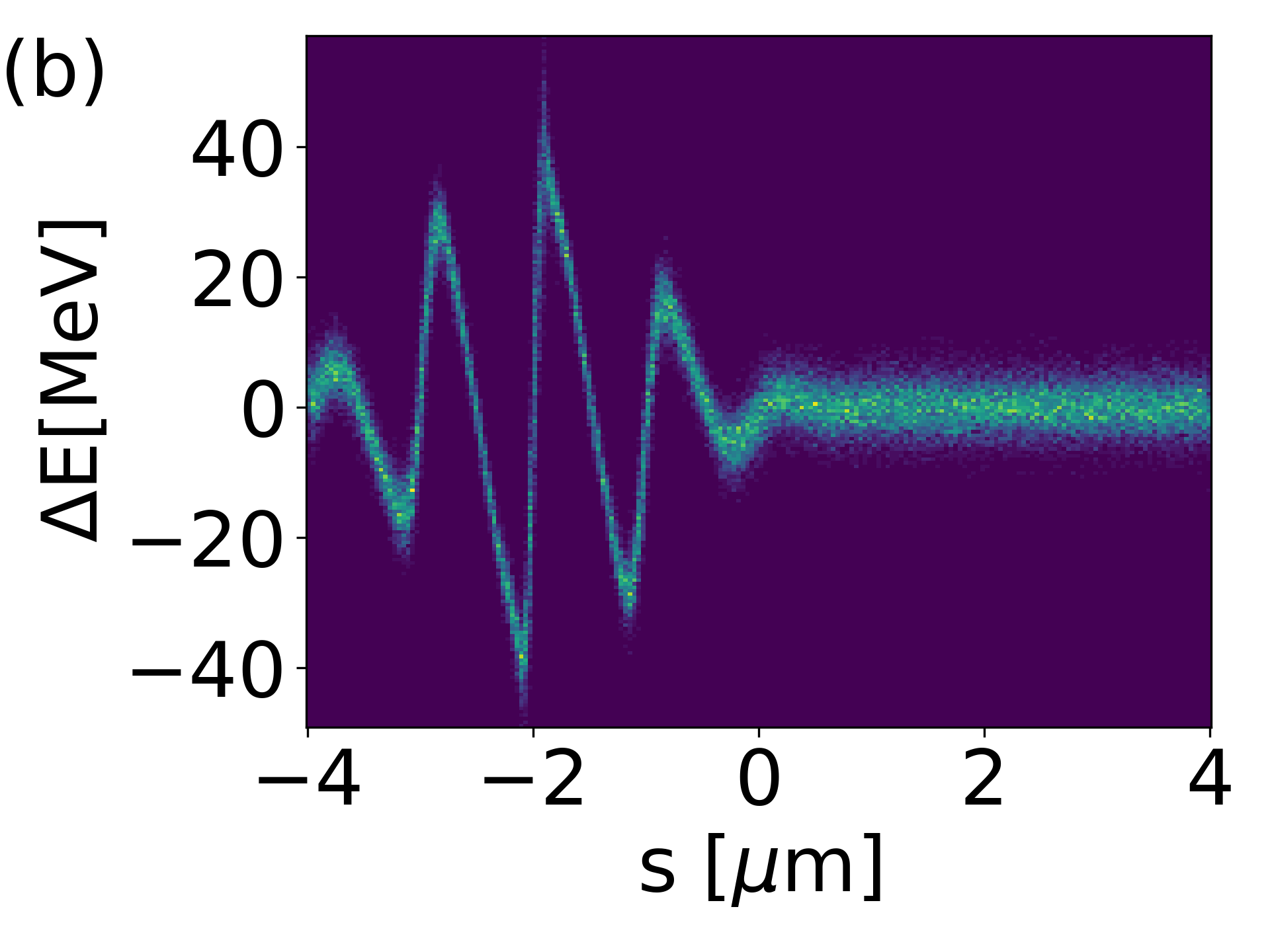}}
	\end{center}
	\caption{The current (a) and energy (b) of the modulated electron beam.}
\end{figure}

To further illustrate the proposed scheme, a detailed example based on the parameters of the European XFEL is studied with the help of Genesis2\cite{reiche1999genesis}  simulations. A 14 GeV electron beam with a normalized emittance of 0.5 mm mrad, a bunch length of 20 fs, and a current flattop profile of 5000 A is assumed here to produce 6 keV FEL pulses. Our simulation study starts with the modulation simulation in \cite{yansimulation}. The electron beam energy modulation is obtained via copropagation of the electron beam with a laser pulse in a wiggler magnet. The electron beam is modulated by a laser with a wavelength of 1030 nm, a pulse energy of 4 mJ and an FWHM pulse duration of 4 fs in the two-periods wiggler with a period of 0.7 m. The tail of the electron beam is thereby modulated in energy. The energy and peak current of the beam after modulation is shown as Figure~\ref{fig:2a} and \ref{fig:2b}. The electron beam with  optical energy modulation is employed in the first part of the  X-ray FEL undulator to generate a light pulse. Then, a SPP is used to impart a helical phase of exp(\textit{il$\phi$}), corresponding to $l=1$, on the pulse. We assumed the intensity of the pulse is unchanged by the SPP. The distance between the SPP and both undulator sections is set to 2 m. 

As indicated by the theoretical analysis, the $l=0$ OAM mode of the SASE pulse is transformed to the $l=1$ OAM mode while the the $l=-1$ mode is transformed to the $l=0$ mode after the filtering. Therefore, the resonance and taper of the first stage undulator need to be optimized to achieve the largest possible ratio (the energy components of different topological charges over the total energy) of the $l=0$ mode while keeping the pulse energy small. Meanwhile, the undulator of the second stage needs to be optimized to achieve the strongest possible amplification of the $l=1$ mode.

\begin{figure}[h]
        \setlength{\abovecaptionskip}{-0.4cm}
        \captionsetup[subfloat]{labelsep=none,format=plain,labelformat=empty}
	\begin{center}
		\subfloat[\label{fig:3a}]{			
			\includegraphics[width=4cm]{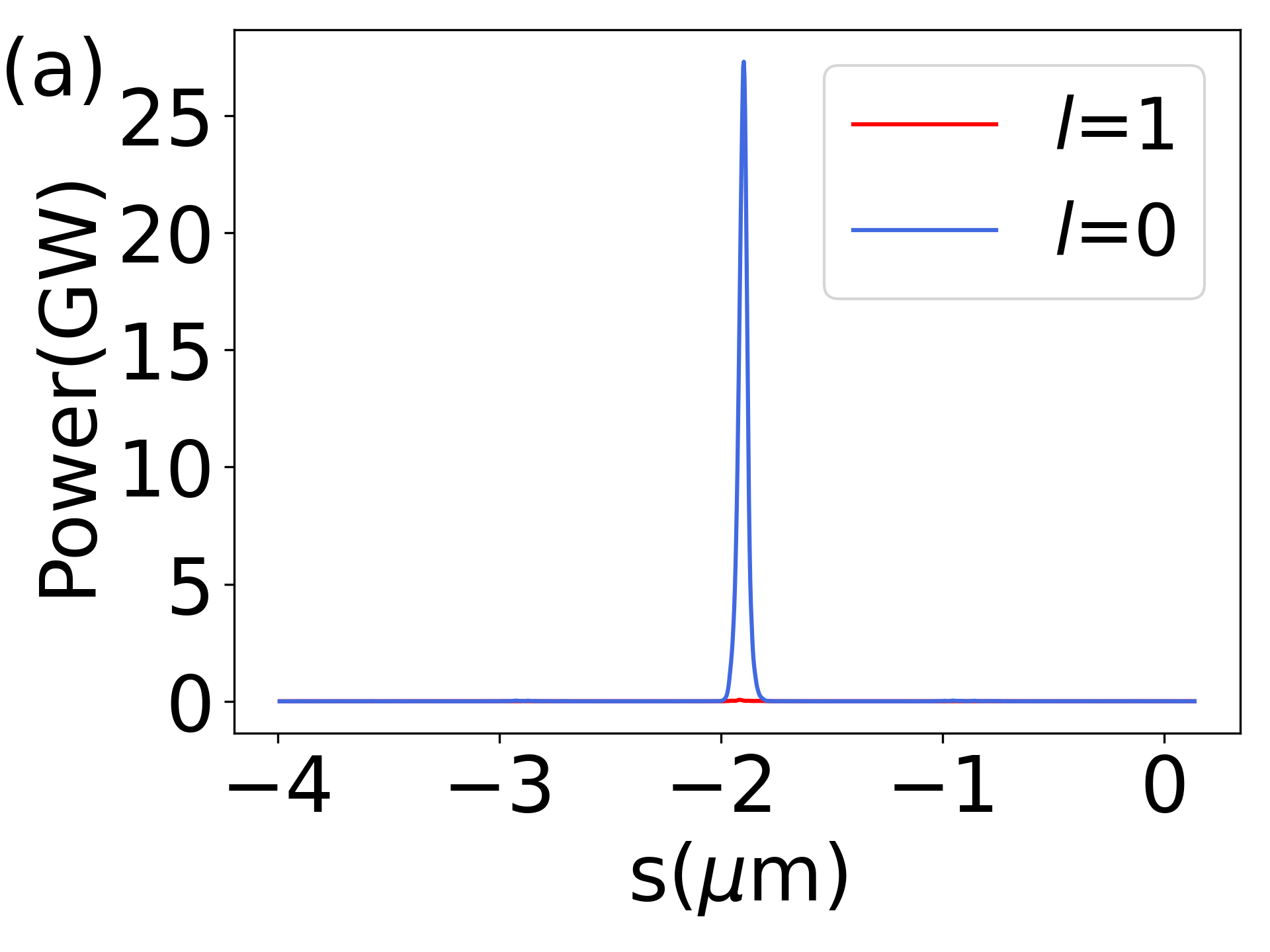}}
		\subfloat[\label{fig:3b}]{			
			\includegraphics[width=4cm]{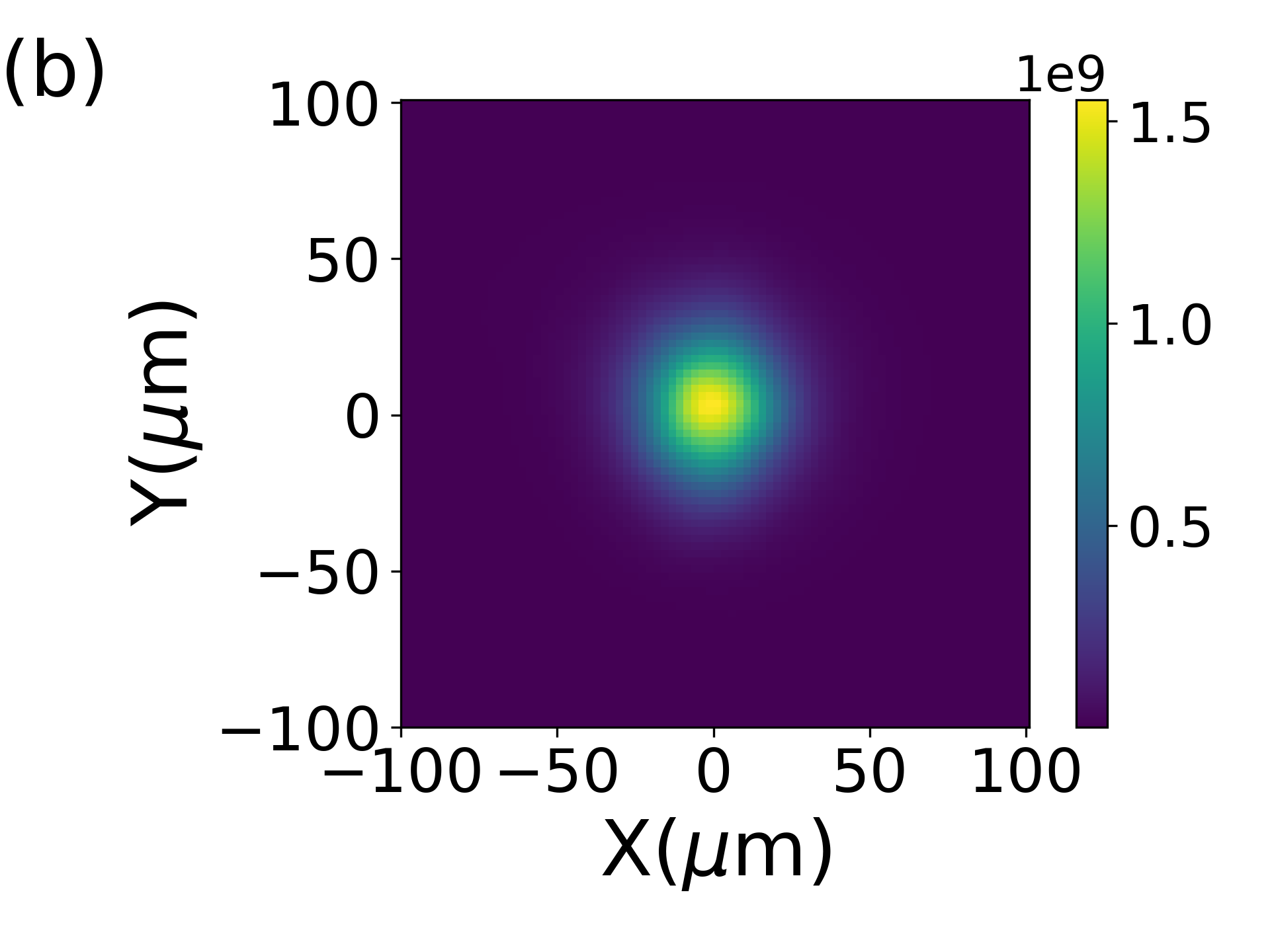}}
	\end{center}
	\caption{Temporal power of FEL pulse (a) and transverse profiles (b) at the end of the first stage.}
\end{figure}

After the modulated electron beam enters the first-stage undulator, it undergoes the process of generating an attosecond pulse. A reverse step-taper shown in Figure~\ref{fig:1b} is applied in the first-stage undulator to match the energy chirp of the energy-modulated portion of the electron beam. The dimensionless undulator parameter K is optimized to increase by 0.0058 in each undulator segment. In this case, an pulse with pulse energy of 4.89 $\mu $J is obtained at the end of the first stage. The mode decomposition of the pulse shows that the relative weight of the \textit{l}=0, \textit{l}=1, and \textit{l}=-1 modes are 94.4\%, 1\%, and 1.7\%, respectively. The temporal power profile and transverse profile of pulse are shown as Figure~\ref{fig:3a} and \ref{fig:3b}. The pulse duration is 122 attosecond at the end of the undulation 8. 

\begin{figure}[h]
        \setlength{\abovecaptionskip}{-0.4cm}
        \captionsetup[subfloat]{labelsep=none,format=plain,labelformat=empty}
	\begin{center}
		\subfloat[\label{fig:4a}]{			
			\includegraphics[width=4cm]{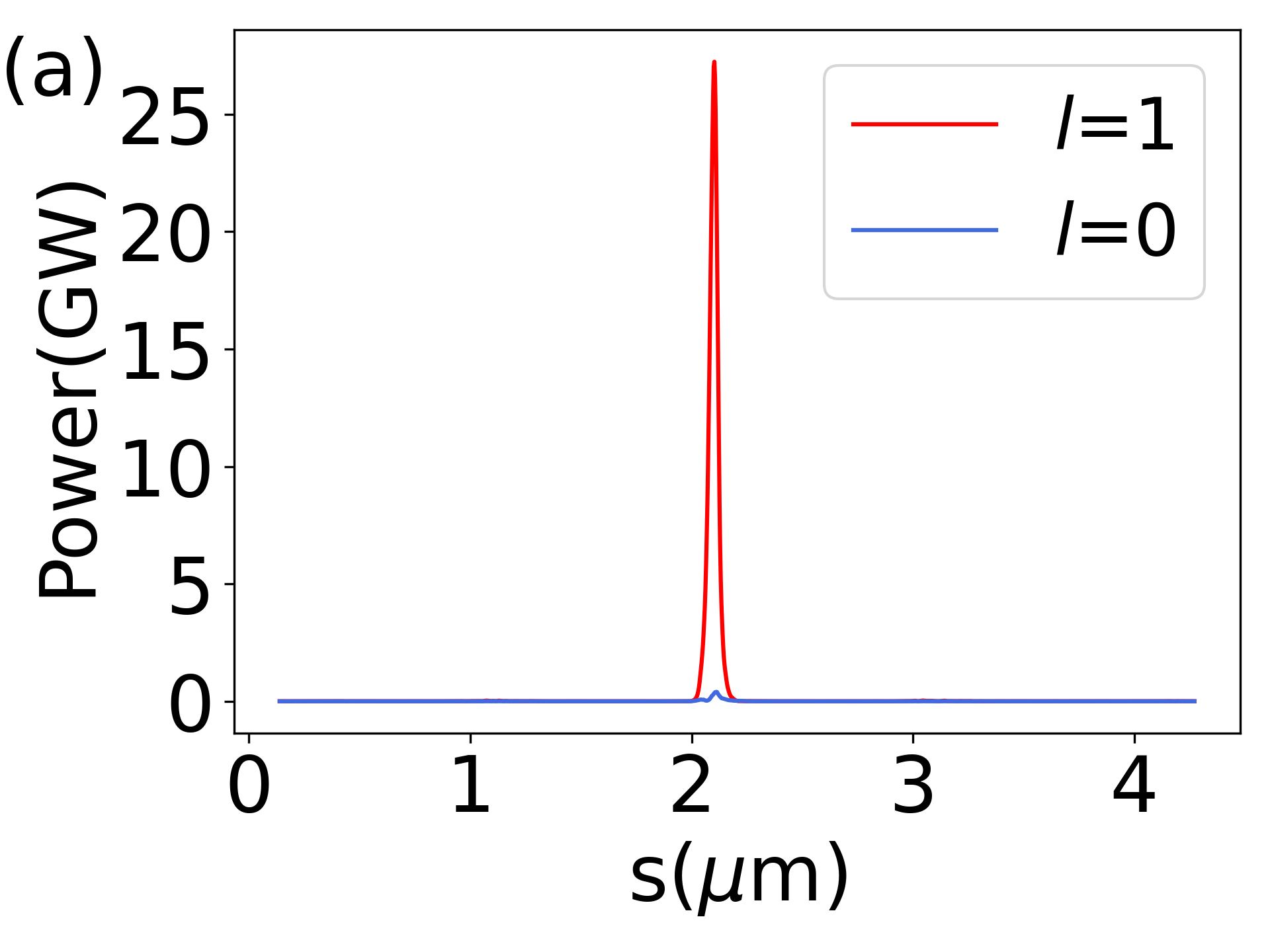}}
		\subfloat[\label{fig:4b}]{			
			\includegraphics[width=4cm]{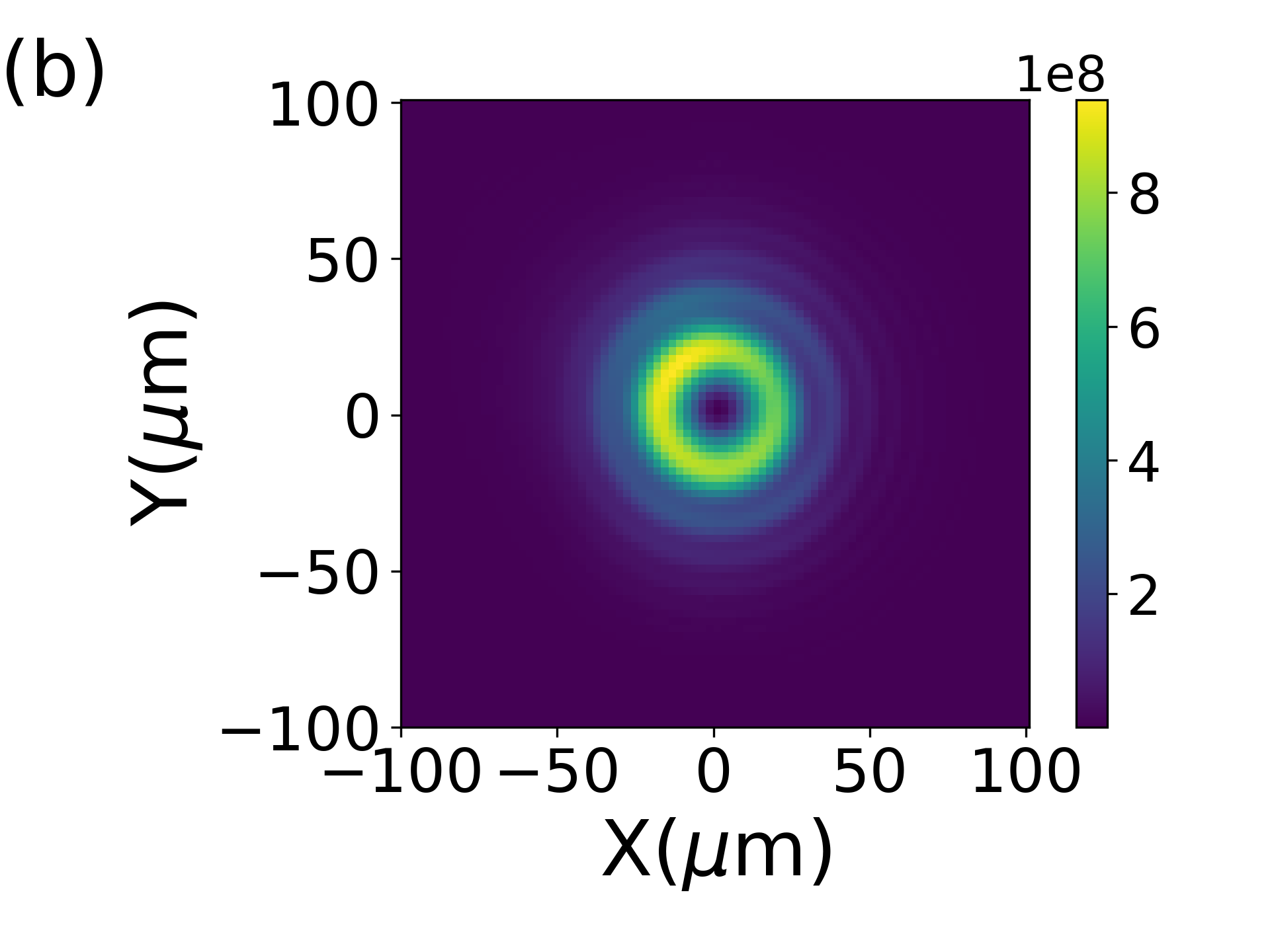}}
		\\
		\subfloat[\label{fig:4c}]{			
			\includegraphics[width=4cm]{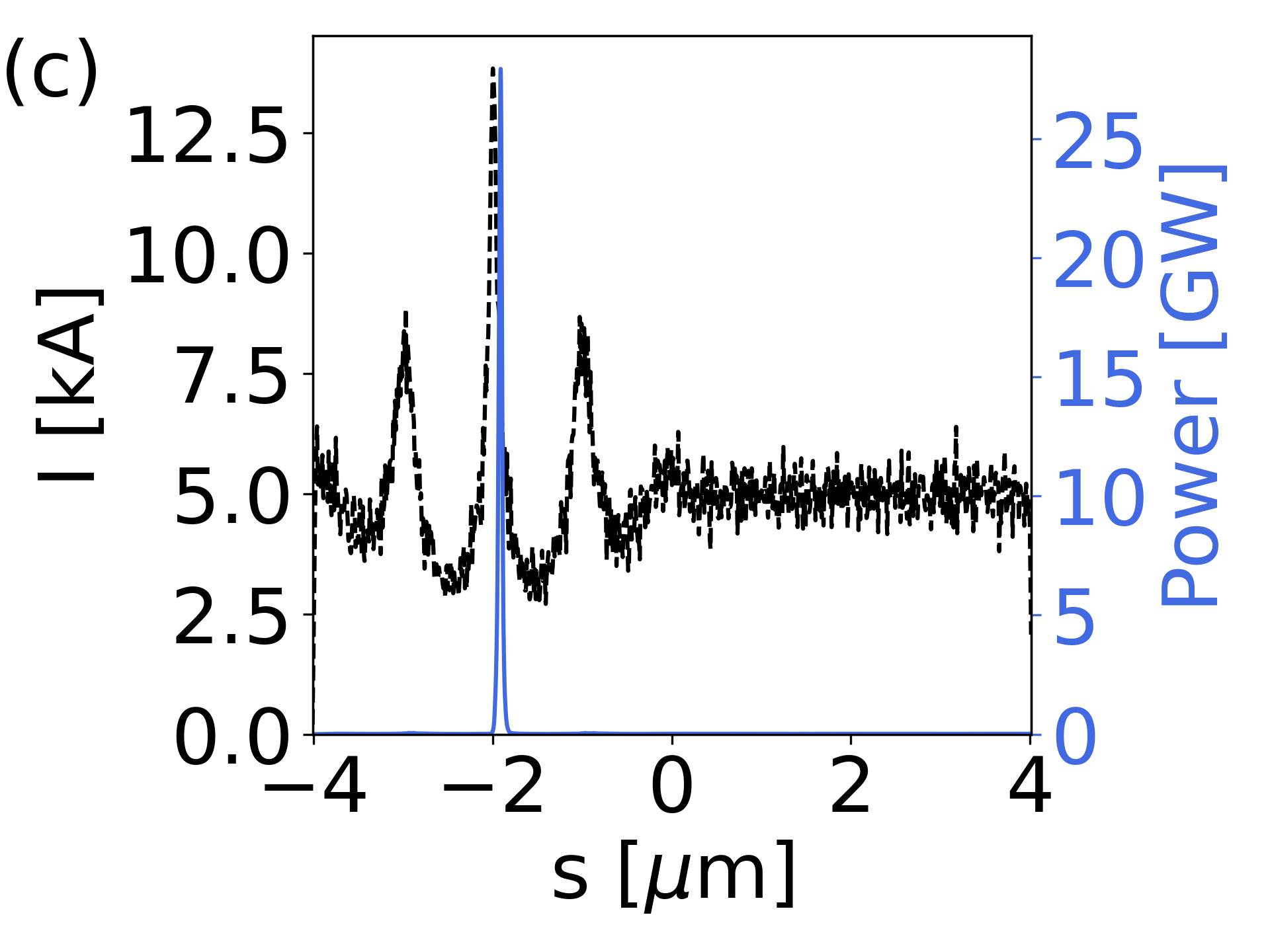}}
		\subfloat[\label{fig:4d}]{			
			\includegraphics[width=4cm]{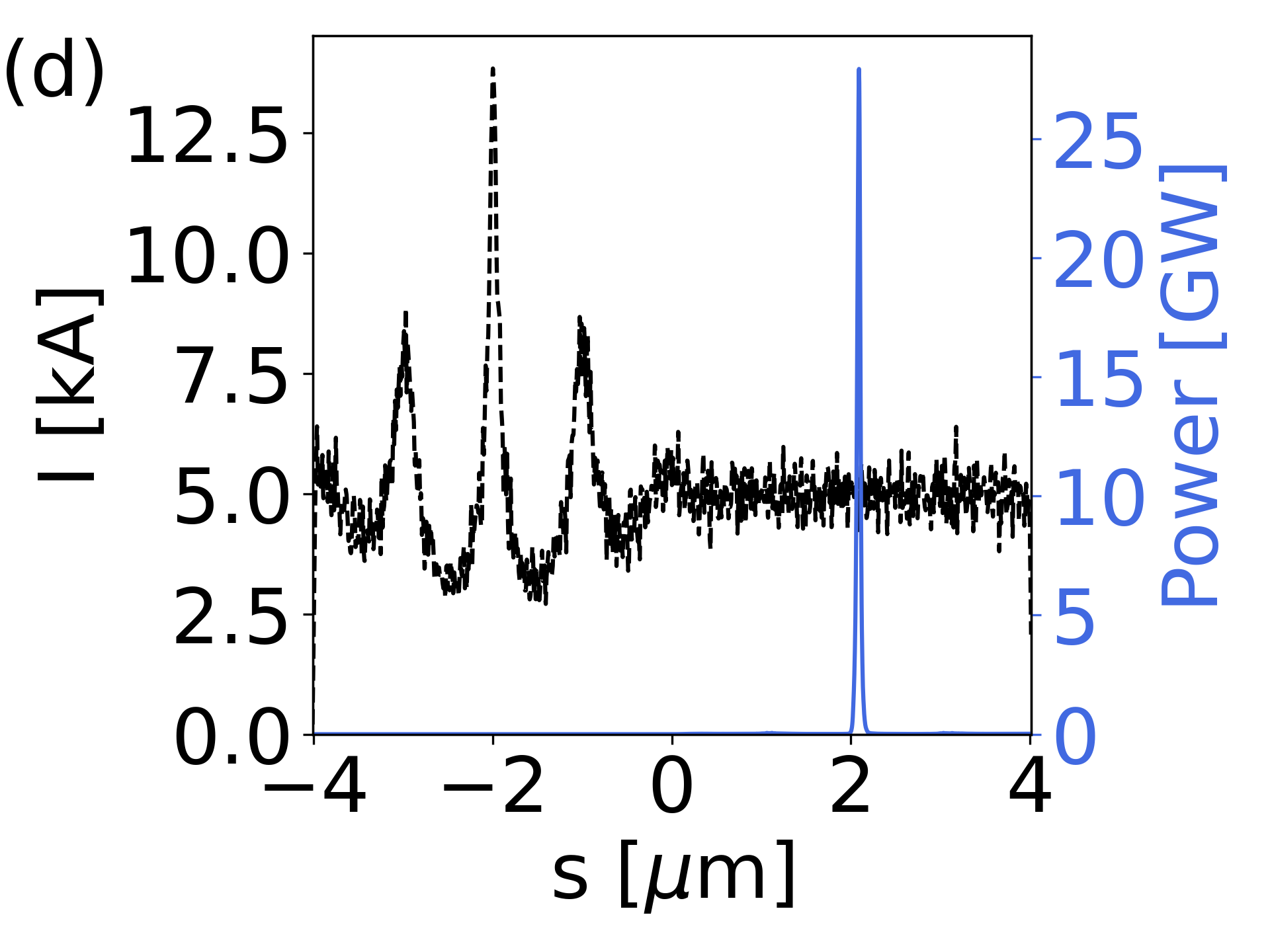}}
	\end{center}
	\caption{Temporal power of the FEL pulse (a) and transverse profiles (b) at the entrance of the second stage decomposed into OAM modes. The position of the pulse relative to the electron beam at (c) the end of the first stage and (d) the entrance of the second stage.}
	\label{fig:first_stage}
\end{figure}

Afterwards, a SPP is employed to introduce a helical phase, exp(\textit{il$\phi$}), corresponding to \textit{l}=1. It is assumed that the intensity of the pulse remains unchanged after passing through the optical element. The distance between the SPP and both undulator sections is set to 2 meters. The FEL pulse propagates through free space, SPP, and free space before reaching the second undulator. At the time of reaching the second stage amplifier, the pulse mainly consist of the \textit{l}=1 mode, with the ratio of 94.4\%, meanwhile the ratio of the \textit{l}=0 mode is 1.7\% as shown in Figure~\ref{fig:4a} and \ref{fig:4b}. To make the laser pulse coincide with the unmodulated part of the electron beam for amplification, set $R_{56}$ to -9 microns. Comparing Figure~\ref{fig:4c} and \ref{fig:4d}, the position of the laser relative to the electron beam is different.

\begin{figure}[h]
        \setlength{\abovecaptionskip}{-0.4cm}
        \captionsetup[subfloat]{labelsep=none,format=plain,labelformat=empty}
	\begin{center}
		\subfloat[\label{fig:5a}]{			
			\includegraphics[width=4cm]{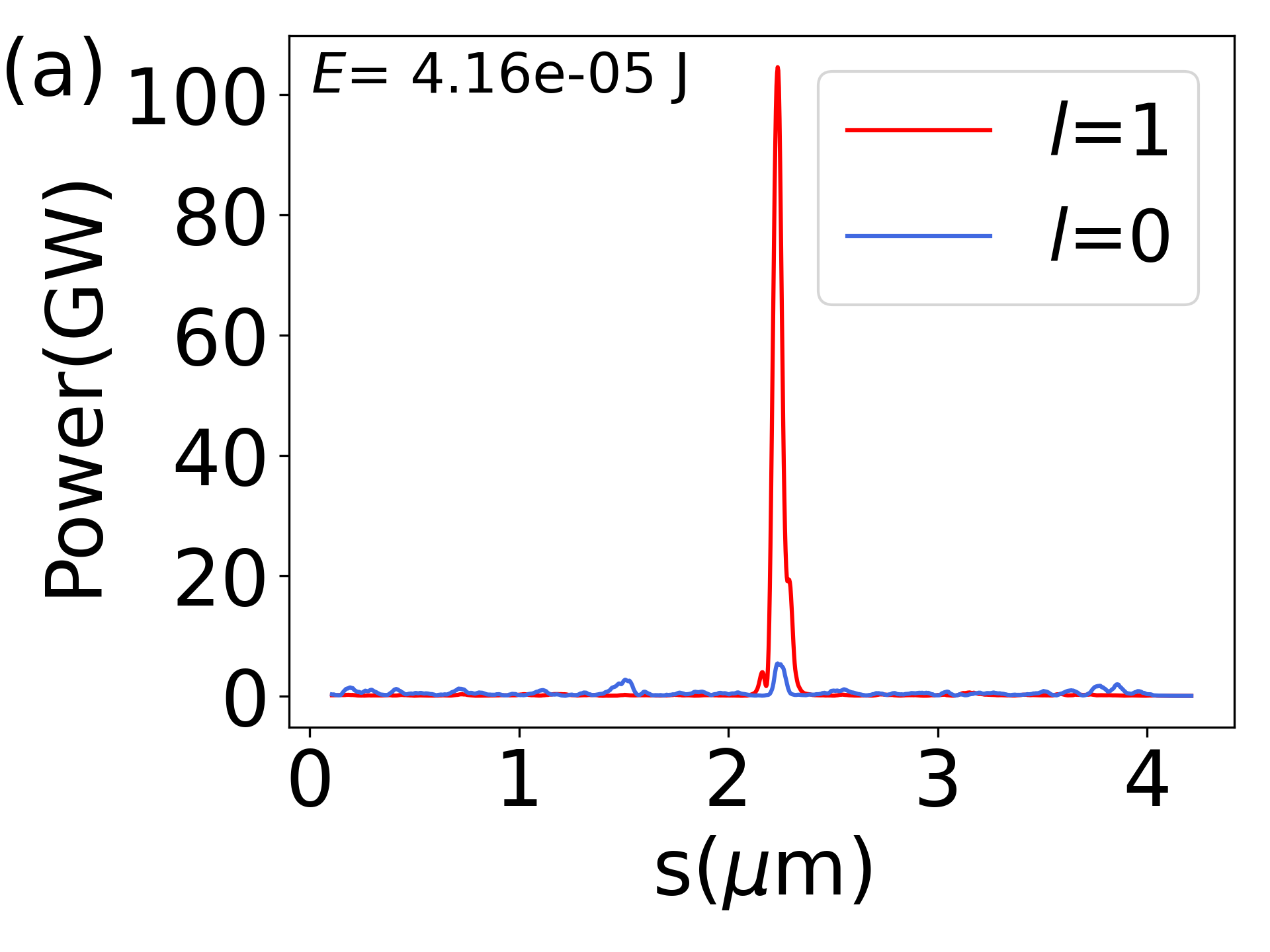}}
		\subfloat[\label{fig:5b}]{			
			\includegraphics[width=4cm]{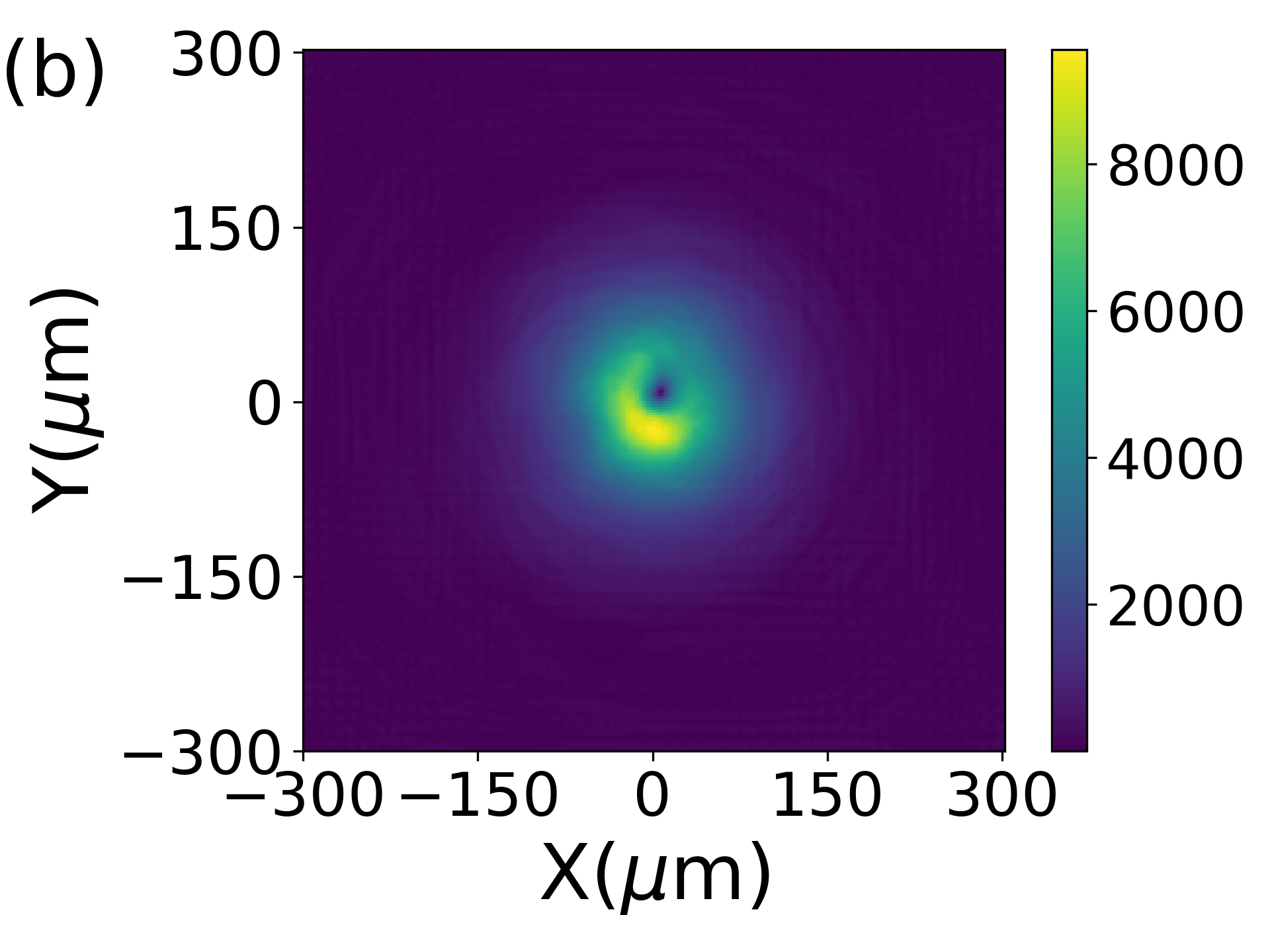}}\\
            \subfloat[\label{fig:5c}]{			
			\includegraphics[width=4cm]{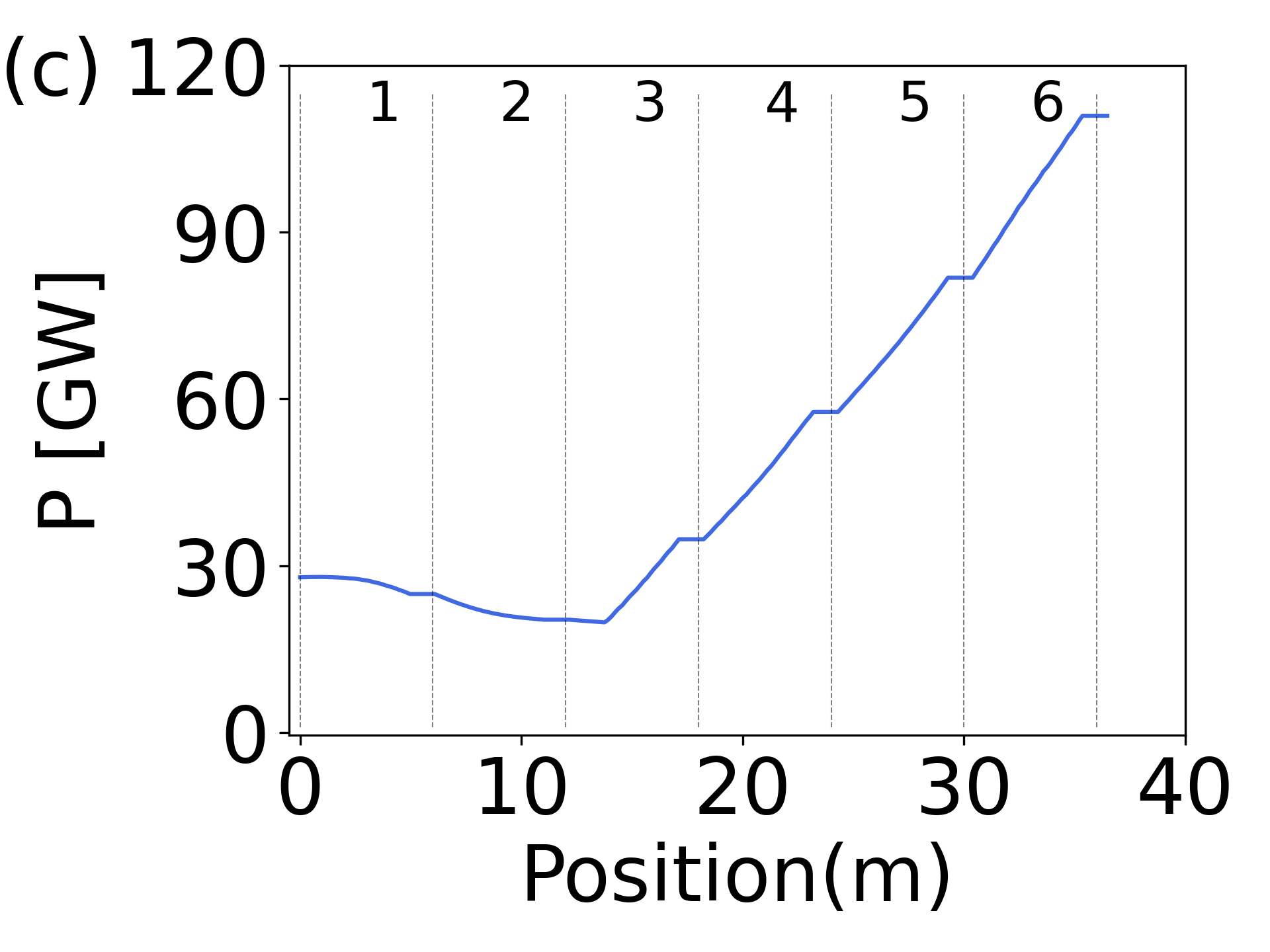}}
		\subfloat[\label{fig:5d}]{			
			\includegraphics[width=4cm]{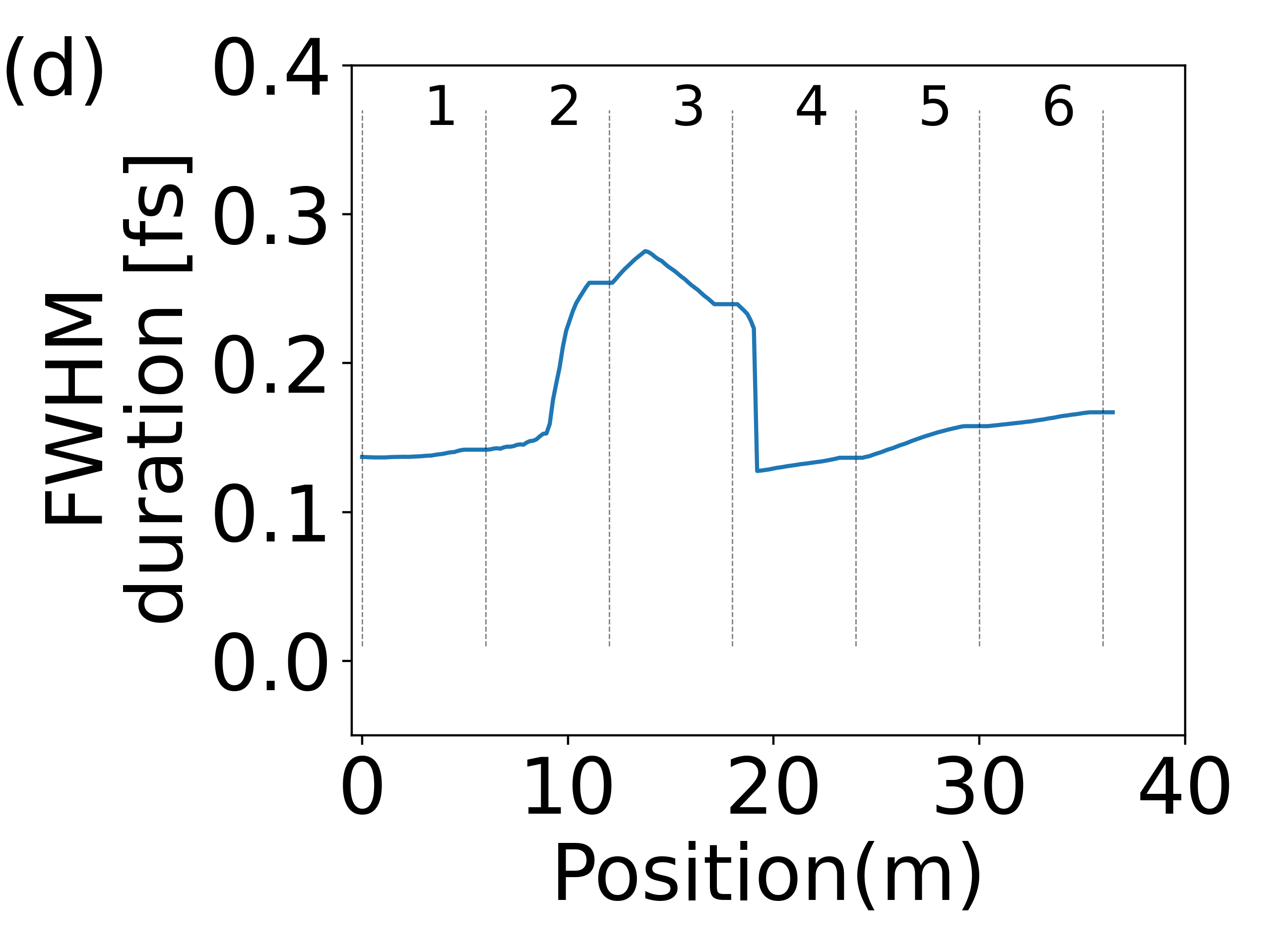}}
	\end{center}
	\caption{Temporal power of the FEL pulse at (a) the end of the second stage. The corresponding transverse profile at peak power is shown in (b). The peak power (c) and the pulse duration (d) changes with the distance of the second stage.}
	\label{fig:one_spike}
\end{figure}

In the second stage, consisting of six undulator segments, the OAM pulse undergoes amplification. The taper of the second undulator is shown in Figure~\ref{fig:1c}. At the exit of the second stage, the pulse energy reaches 41.6 $\mu $J as shown in Figure~\ref{fig:5c}. Figure~\ref{fig:5b} displays the transverse phase distribution at the position of maximum power in the pulse at the second stage's end. Figure~\ref{fig:5a} shows that the pulse duration is 164 attosecond and the peak power of \textit{l}=1 is 102 GW as shown in Figure~\ref{fig:5d}. Since there will be growth in places other than the spike in the second stage, the power of \textit{l}=1 accounts for 62\% of the total pulse. After seven undulator segments, the total ratio of \textit{l}=1 will drop below 50\%.

\section{CONCLUSION}

We present a novel approach for producing high-purity X-ray OAM pulses characterized by both high peak power over 100 GW and attosecond-range  pulse duration. The generation and amplification of attosecond OAM pulses are demonstrated by simulations.

\section{ACKNOWLEDGEMENTS}
This work was supported by the CAS Project for Young Scientists in Basic Research (Grant No. YSBR-042), the National Natural Science Foundation of China (Grant Nos. 12125508 and 11935020), the Program of Shanghai Academic/Technology Research Leader (Grant No. 21XD1404100), and the Shanghai Pilot Program for Basic Research of the Chinese Academy of Sciences, Shanghai Branch (Grant No. JCYJ-SHFY-2021-010).

%
%
\bibliographystyle{abbrv}
\bibliography{TUP146-FRA}
%
%


\end{document}